\documentstyle[12pt,epsf]{article}
\begin{document}
\setlength\textheight{8.75in}
\newcommand{\be}{\begin{equation}}
\newcommand{\ee}{\end{equation}}
\title{\bf Spherically symmetric Yang-Mills solutions  in
a 5-dimensional (Anti-) de Sitter space-time}
\author{{\large Betti Hartmann\footnote{ betti.hartmann@durham.ac.uk}}\\
{\small Department of Mathematical Sciences,
University of Durham,}\\
{\small Durham DH1 \ 3LE, United Kingdom}\\
{ }\\
{\large Yves Brihaye\footnote{ yves.brihaye@umh.ac.be}} \\
{\small Facult\'e des Sciences, Universit\'e de Mons-Hainaut, }\\
{\small B-7000 Mons, Belgium }\\
{ } \\
 {\large Bruno Bertrand} \\
{\small Facult\'e des Sciences, Universit\'e de Mons-Hainaut, }\\
{\small B-7000 Mons, Belgium }\\  }

\date{\today}

\maketitle
\thispagestyle{empty}

\begin{abstract}
We consider an Einstein-Yang-Mills Lagrangian
in a five dimensional space-time including a cosmological 
constant. Assuming all fields to be independent of the
extra coordinate, a dimensional reduction
leads to an effective
(3+1)-dimensional Einstein-Yang-Mills-Higgs-dilaton
 model where the cosmological constant induces a Liouville
 potential in the dilaton field.
We construct spherically symmetric 
solutions analytically in specific limits and
study the generic solutions for vanishing dilaton coupling numerically.
We find that in this latter case the solutions bifurcate with the
branch of (Anti-) de Sitter-Reissner-Nordstr\"om ((A)dSRN) solutions.

\end{abstract}
\medskip
\medskip
\newpage
\section{Introduction}
The scalar dilaton field arised as companion of the metric tensor
in (super)string theories and is associated with the scale invariance of these
theories \cite{maeda}. Thus it is interesting to study classical field theory
solutions coupled to a dilaton. In most studies, the
dilaton was assumed to be massless while, however, from the viewpoint
of a realistic theory the dilaton should be massive
in order to avoid long-range scalar forces. In \cite{chm} a dilaton
potenial of Liouville type was introduced to take into account
the effects of a specific symmetry-breaking mechanism which gives
mass to the dilaton. This type of potential has a constant prefactor
which in the limit of vanishing dilaton coupling reduces to a cosmological
constant. It was found that there exist no asymptotically flat/ de Sitter/
Anti-de Sitter solutions for non-vanishing potential \cite{polet}.
Rotating generalisations of the black hole solutions found in \cite{chm}
have been constructed in \cite{mitra}.

Volkov argued recently \cite{volkov} that if $\frac{\partial}{\partial x_4}$
is a symmetry of the Einstein-Yang-Mills (EYM) system in ($4+1$) dimensions, where 
$x_4$ is the coordinate associated with the $5$th dimensions, than the
($4+1$)-dimensional EYM system reduces effectively to a ($3+1$)-dimensional EYMHD system
with a specific coupling between the dilaton field and the Higgs field.
The generalisation of this ($3+1$)-
dimensional EYMHD model was consequently studied in \cite{bh}.

In this paper, we study spherically symmetric solutions
of the ($3+1$)-dimensional EYMHD model deduced from the 
($4+1$)-dimensional EYM system including a cosmological constant. The dimensional
reduction then leads to a Liouville-type potential in the $(3+1)$-dimensional
model. In Section 2, we present both the five-dimensional model and the
from this deduced and then generalised $(3+1)$-dimensional EYMHD model. In Section 3, we discuss
the solutions for the case of vanishing dilaton coupling, especially, we present
our numerical results for the generic solutions in this case. In section 4, we discuss
possible solutions for the generic case of non-vanishing dilaton coupling. 
Our conclusions are presented in Section 5.

\section{The model}

We start with the Einstein-Yang-Mills Lagrangian
in five dimensions including a cosmological constant
given by:

\begin{equation}
  S = \int \Biggl(
    \frac{1}{16 \pi G_5}( R  - 2  \Lambda_{(5)})
  - \frac{1}{4 \tilde{e}^2}F^a_{M N}F^{a M N}
  \Biggr) \sqrt{g^{(5)}} d^5 x
\end{equation}
with the SU(2) Yang-Mills field strength
$F^a_{M N} = \partial_M A^a_N -
 \partial_N A^a_M + \epsilon_{a b c}  A^b_M A^c_N$
, the gauge index
 $a=1,2,3$  and the space-time index
 $M=0,1,2,3,4$. $G_5$ and $\tilde{e}$ denote
respectively the 5-dimensional Newton's constant and the coupling
constant of the gauge field theory. $G_5$ is related to the
5-dimensional Planck scale $M_{Pl(5)}$ by $G_5=M^{-3}_{Pl(5)}$.
$\Lambda_{(5)}$ is the 5-dimensional
cosmological constant.

If both the matter functions and the metric functions
are independent on $x_4$, the $5$-dimensional fields can be
parametrized as follows  \cite{volkov}:
\begin{equation}
g^{(5)}_{MN}dx^M dx^N = e^{-\zeta}g^{(4)}_{\mu\nu}dx^{\mu}dx^{\nu}
+e^{2\zeta} (dx^4)^2 \ , \ \mu , \nu=0, 1, 2, 3
\end{equation}
and
\begin{equation}
A_M^{a}dx^M=A_{\mu}^a dx^{\mu}+\Phi^a dx^4  \  , \
\end{equation}
where $g^{(4)}$ is the $4$-dimensional metric tensor and $\zeta$ plays the
role of the dilaton.

In \cite{volkov} it was shown that for $\Lambda_{(5)}=0$ the classical equations
are equivalent to those of a four-dimensional
Einstein-Yang-Mills-Higgs dilaton theory.
In this paper, we consider the case with a cosmological constant. We then choose
the generalised $(3+1)$-dimensional action to be:
\begin{equation}
S=S_{G}+S_{M}=\int L_{G}\sqrt{-g^{(4)}}d^{4}x+
\int L_{M}\sqrt{-g^{(4)}}d^{4}x
\ . \label{action} \end{equation}
with the gravity Lagrangian:
\begin{equation}
L_{G}=\frac{1}{16\pi G}R 
\ ,\end{equation}
and $G$ denoting the 4-dimensional Newton`s constant. The matter
Lagrangian $L_M$ reads:
\begin{equation}
L_{M}=-\frac{1}{4} e^{2\kappa\Psi}F_{\mu\nu}^{a}F^{\mu\nu,a}
-\frac{1}{2}\partial_{\mu}\Psi\partial^{\mu}\Psi
-\frac{1}{2}e^{-4\kappa\Psi}D_{\mu}\Phi^{a}
D^{\mu}\Phi^{a}-e^{-2\kappa\Psi}V(\Phi^{a})
- \frac{\tilde{\Lambda}}{2}e^{-2\kappa\Psi}
\ , \label{lag}  \end{equation} 
with the Higgs potential
\begin{equation}
V(\Phi^{a})=\frac{\lambda}{4}(\Phi^{a}\Phi^{a}-v^2)^2 
\ , \end{equation}
the non-abelian field strength tensor 
\begin{equation}
F_{\mu\nu}^{a}=\partial_{\mu}A_{\nu}^{a}-\partial_{\nu}A_{\mu}^{a}+
e\varepsilon_{abc}A_{\mu}^{b}A_{\nu}^{c}
\ , \end{equation}
and the covariant derivative of the Higgs field in the adjoint
representation
\begin{equation}
D_{\mu}\Phi^{a}=\partial_{\mu}\Phi^{a}+
e\varepsilon_{abc}A_{\mu}^{b}\Phi^{c}
\ . \end{equation}
 The gauge field coupling constant is denoted $e$,
$\lambda$ is the Higgs field coupling constant
and $v$ the vacuum expectation value of the Higgs field.

Note that we have introduced a coupling $\kappa$ specific to the dilaton field
 by setting $\zeta=2\kappa\Psi$.
 This will allow to study the influence of the dilaton
systematically. We remark that the $5$-dimensional cosmological constant has through dimensional
reduction led to
 a Liouville potential in the dilaton field with coupling constant
 $\tilde{\Lambda}$. For $\kappa=0$, $\tilde{\Lambda}$ is proportional
 to the four-dimensional
 cosmological constant.

\subsection{The Ansatz}

For the metric the spherically symmetric Ansatz
in Schwarzschild-like coordinates reads \cite{bfm,weinberg}:
\begin{equation}
ds^{2}=g^{(4)}_{\mu\nu}dx^{\mu}dx^{\nu}=
-A^{2}(r)N(r)dt^2+N^{-1}(r)dr^2+r^2 d\theta^2+r^2\sin^2\theta
d^2\varphi
\label{metric}
\end{equation}
with
\begin{equation}
N(r)=1-\frac{2m(r)}{r}
\ . \end{equation}
In these coordinates, $m(\infty)$ denotes the (dimensionful) mass of
the field configuration.\

For the gauge and Higgs fields, we use the purely magnetic hedgehog ansatz
\cite{thooft}
\begin{equation}
{A_r}^a={A_t}^a=0
\ , \end{equation}
\begin{equation}
{A_{\theta}}^a= \frac{1-K(r)}{e} {e_{\varphi}}^a
\ , \ \ \ \
{A_{\varphi}}^a=- \frac{1-K(r)}{e}\sin\theta {e_{\theta}}^a
\ , \end{equation}
\begin{equation}
{\Phi}^a=v H(r) {e_r}^a
\ . \end{equation}
The dilaton is a scalar field depending only on $r$
\begin{equation}
\Psi=\Psi(r)
\ . \end{equation}
Inserting the Ansatz into the Lagrangian and varying with respect
to the matter fields yields the Euler-Lagrange equations,
while variation with respect to the metric yields the Einstein
equations.\

\subsection{Classical field equations}

With the introduction of  dimensionless coordinates and fields
\begin{equation}
x=evr \ , \ \ \mu=evm \ ,\ \ \phi=\frac{\Phi}{v}\ , \ \
\psi=\frac{\Psi}{v}
\label{scale}
 \end{equation}
the Lagrangian and the resulting set of differential equations
depend on the following
coup\-ling constants:
\begin{equation}
\alpha =\sqrt{G}v =\frac{M_W}{eM_{\rm Pl}} \ , \ \
\beta=
 \frac{\sqrt{\lambda}}{e} = \frac{M_H}{\sqrt{2}M_W} \ , \ \
\gamma =\kappa v =\frac{\kappa M_W}{e}
 \label{coupling} \ , \ \ \Lambda=2\alpha^2 \tilde{\Lambda} \ ,
\end{equation}
where $M_W=e v$, $M_H= \sqrt{2\lambda} v$ and $M_{\rm Pl}=1/\sqrt{G}$. With
the rescalings (\ref{scale}) and (\ref{coupling}), the dimensionless
mass of the solution is given by $\frac{\mu(\infty)}{\alpha^2}$.
Note that we have rescaled the cosmological constant in order to obtain
the equations of a conventional (3+1)-dimensional Einstein-Yang-Mills-Higgs
model including a cosmological constant in the limit of vanishing dilaton 
coupling. 

With (\ref{scale}) and (\ref{coupling}) the Euler-Lagrange equations read:
\begin{equation}
(e^{2\gamma\psi}ANK')'=A(e^{2\gamma\psi}\frac{K(K^2-1)}{x^2}+
e^{-4\gamma\psi}H^2
K)
\ , \label{dgl1} \end{equation}
\begin{equation}
(e^{-4\gamma\psi}x^2 ANH')'=AH(2e^{-4\gamma\psi}K^2+ \beta^2 x^2
e^{-2\gamma\psi}(H^2-1))
\ , \label{dgl2}
\end{equation}
\begin{eqnarray}
(x^2 A N\psi')' &=& 2\gamma A [e^{2\gamma\psi}(N(K')^2+\frac{(K^2-1)^2}{2
x^2}) -\frac{\Lambda}{4\alpha^2} x^2 e^{-2\gamma\psi}
\nonumber \\
&-& e^{-2\gamma\psi}\frac{\beta^2 x^2}{4}(H^2-1)^2-2
e^{-4\gamma\psi}(\frac{1}{2}
N (H')^2 x^2+H^2 K^2) ]
\ , \label{dgl3} \end{eqnarray}
where the prime denotes the derivative with respect to $x$,
while we use the following combination of the Einstein equations
\begin{equation}
G_{tt}=2\alpha^2 T_{tt}=-2\alpha^2 A^2 N L_{M}
\ , \end{equation}
\begin{equation}
g^{xx}G_{xx}-g^{tt}G_{tt}=-4\alpha^2 N \frac{\partial L_{M}}{\partial N}
  \end{equation}
to obtain two differential equations for the two metric functions:
$$
\mu ' = \alpha^2 \left(
e^{2\gamma\psi}N(K')^2 + \frac{1}{2}N x^2(H')^2
e^{-4\gamma\psi}+
\frac{1}{2x^2}(K^2-1)^{2} e^{2\gamma\psi}+K^2 H^2 e^{-4\gamma\psi}\right.
$$
\begin{equation}
 +  \left. \frac{\beta^{2}}{4}x^2
(H^2-1)^2 e^{-2\gamma\psi}+\frac{1}{2}Nx^{2}(\psi ')^2\right)
+ \frac{\Lambda}{4} x^2  e^{-2\gamma\psi} 
  \  , \label{dgl4}
\end{equation}
\begin{equation}
A'=\alpha^2 x A \left(\frac{2(K')^2}{x^2}e^{2\gamma\psi}+
e^{-4\gamma\psi}(H')^2+(\psi ')^2
\right)
\ . \label{dgl5} \end{equation}

Note that the equations of the original five dimensional theory are recovered
by using the following specific choice of the coupling constants:
\begin{equation}
\label{5dlimit}
 \alpha^2 = 3 \gamma^2 \ \ , \ \
  \Lambda = \Lambda_{(5)}  \ \ , \ \
  \beta = 0 \ \ .
\end{equation}

The case $\Lambda=0$ was previously studied in \cite{volkov,bh}.
If in addition $\gamma=0$, the equations of the Einstein-Yang-Mills-Higgs
equations are recovered \cite{bfm,weinberg}. Choosing
$\Lambda=\alpha=0$ (assuming $\Lambda/\alpha^2=0$ as well), the model reduces to the Yang-Mills-Higgs-dilaton
system studied in \cite{forgacs1}.

\section{Spherically symmetric solutions for $\gamma=0$ }

We will first discuss the solutions in the case
$\gamma = 0$.  The equation of the dilaton field
can then be decoupled and $\psi(x)\equiv 0$.
We will study solutions of this system
which are regular at the origin
this implies the following conditions
\begin{equation}
K(0)=1 \ , \ \ H(0)=0 \ , \ \ \mu(0)=0
\ . \label{bc1}
\end{equation}
Finiteness of the ADM mass requires that the fields approach
particular values  asymptotically, namely:
 \begin{equation}
K(\infty)=0 \ , \ \ H(\infty)=1 \ , \  \ A(\infty)=1
\ . \label{bc2} \end{equation}

For $\Lambda > 0$ the metric function $N(x)$ has a  zero at a finite
value of $x$, say $x=x_c$. This is the so-called ``cosmological horizon''.
The value $x_c$ depends on the actual values of the coupling constants.

\subsection{(Anti-) De Sitter-Reissner-Nordstr\"om ((A)dSRN) solutions}
Setting $\gamma = 0$
the system admits embedded abelian solutions,
the so-called (Anti-) de Sitter-Reissner-Nordstr\"om
solutions: 
\begin{equation}
K(x) = 0 \ \ , \ \ H(x) = 1 \ \ , \ \ \psi(x) = 0 \ \ , \ \ A(x) = 1 \ \ , \ \
N(x)= 1-\frac{1}{6} \Lambda x^2 -\frac{2\mu_{\infty}}{x}+\frac{\alpha^2}{x^2} \ .
\end{equation}
The metric function $N(x)$ has a physical singularity at the origin $x=0$ which
is evident from the Kretschmann scalar $K=R_{\alpha\beta\gamma\delta}
R^{\alpha\beta\gamma\delta}$:
\begin{equation}
K=\frac{2}{3x^8}\left(\Lambda^2x^8+72 \mu_{\infty}^2x^2-144 \mu_{\infty}
\alpha^2 x+84 \alpha^2\right) \ .
\end{equation}
Depending on the choice of the sign of the cosmological constant, up to 
$4$ zeros of $N(x)$ can exist. $3$ of the $4$ zeros correspond to horizons
since the first zero has always negative value and thus has no physical meaning.
The two inner horizons $x_-$, $x_+$ with $x_- \le x_+$ correspond
to the well known Cauchy, respectively event horizon of the
Reissner-Nordstr\"om solution, while the third outer horizon
$x_c > x_+$ exists only for positive cosmological constant.

Extremal black hole solutions - like in the asymptotically flat space - are
possible. Then, we have $x_-=x_+=x_h$ with $N(x_h)=N'|_{x=x_h}=0$.
This leads to the equation:
\begin{equation}
\Lambda x_h^4 -2x_h^2+2\alpha^2=0 \ .
\end{equation}
This is solved by:
\begin{equation}
\label{xhds}
x_{h/c}=\frac{1}{\sqrt{\Lambda}}\sqrt{1\pm\sqrt{1-2\alpha^2\Lambda}} \ \
\ \ \ {\rm for} \ \ \  \frac{1}{2\alpha^2} \ge \Lambda > 0
\end{equation}
 and
 \begin{equation}
x_{h}=\frac{1}{\sqrt{|\Lambda|}}\sqrt{-1+\sqrt{1-2\alpha^2\Lambda}} \ \
\ \ \ {\rm for} \ \ \ \Lambda < 0 \ .
\end{equation}
For $\Lambda > 0$, the solution with the plus sign is
the outer, cosmological horizon $x_c$, while the inner, event horizon $x_h$ is
the solution with the minus sign. Obviously, the appearance of horizons
in dS space is restricted by $\alpha^2 \le \frac{1}{2\Lambda}$.
The corresponding mass of the extremal solution is given by:
\begin{equation}
\mu_{\infty}=\frac{2}{3}\frac{\alpha^2}{x_h}+\frac{x_h}{3} \ .
\end{equation}
Apparently, the $\Lambda=0$ limit is ill-defined. However,
for $0 < \Lambda \ll 1$ we find
\begin{equation}
\mu_{\infty} = \alpha - \frac{\alpha^3}{9} \Lambda + O(\Lambda^2) \ \ , \ \
x_h = \alpha + \frac{\alpha^3}{3} \Lambda  +  O(\Lambda^2)
\end{equation}
which for $\Lambda\rightarrow 0$ obviously leads to the corresponding
values of the well-known
asymptotically flat Reissner-Nordstr\"om solution.

 \subsection{de Sitter (dS) gravitating monopoles}  
 Since gravitating monopoles in Anti-de Sitter space have been studied 
 previously \cite{adsmono}, we concentrate here on monopoles in
 de Sitter space. To our knowledge, these type of solutions have not been
 studied previously.
 
 In the absence of a cosmological constant, 
 the flat space magnetic monopole \cite{thooft}
 is deformed by gravity and exist up to a critical value of $\alpha=\alpha_{cr}$
 where the solution bifurcates with the branch of extremal Reissner-Nordstr\"om
  solutions \cite{bfm}. For instance in the BPS limit ($\beta = 0$)
  the gravitating monopole bifurcates with this branch
  at $\alpha_{cr} \approx 1.386$.

 Now analysing the equations in the presence of a cosmological
 constant, we were able to construct dS-gravitating  monopoles.
 They are characterised by a cosmological horizon at $x = x_c$
 with $N(x=x_c)=0$. The behaviour of the function $N(x)$ is illustrated in Fig. 1
 for $\alpha = 0.8$ and  different values of $\Lambda$.
 We find that $x_c$ is decreasing with the increase of
 $\Lambda$: $x_c \sim 108$ for
 $\Lambda \sim 0.0005$ and
 $x_c \sim 77$ for $\Lambda = 0.001$.  
 As is obvious from the figure, the solutions have a local minimum
 at some value of the radial coordinate $x=x_{min}(\Lambda)$.

 The main aim of this study was to determine the domain of 
 coupling constants in which dS-gravitating monopoles exist.
 Fixing $\beta$ and $\Lambda$
 our analysis demonstrates that dS-gravitating monopoles bifurcate
 with the branch of extremal dSRN solutions described in the previous
 section at a critical value
 of $\alpha$.    Since we limited our analysis to small values
 of $\Lambda$ the critical value of $\alpha$ where the bifurcation
 occurs hardly differs from the corresponding one in the asymptotically
 flat case.
 
 The way how the extremal dSRN solution  is approached is illustrated in 
 Fig. 2
 for $\Lambda = 0.001$ and $\beta=0.1$. This clearly shows that the  value of the local minimum
 of the function $N(x)$ decreases while $\alpha$ increases. 
 We find that solutions exist up to a maximal value of the gravitational
 coupling $\alpha=\alpha_{max}\approx 1.382$. There 
 another branch of non abelian
   solutions exist which bifurcates with
 the  branch of dSRN solutions at a critical value of $\alpha=\alpha_{cr}\approx
 1.378$. At this point, 
 a degenerate horizon forms at $x=x_h$. The critical solution can be described
 by the dS-RN solution with horizons (\ref{xhds}) for $x \ge x_h$, while
 for $x_h > x \ge 0$, it is non-singular and non-trivial. Compared to the case
 $\Lambda=0$ \cite{bfm}, the values of $\alpha_{max}$ and $\alpha_{cr}$ are
 smaller when $\Lambda > 0$. Moreover, 
 the interval of $\alpha$ on which two solutions exist decreases. 
 This can be related to the increased cosmological expansion for $\Lambda > 0$.
 
\section{Spherically symmetric solutions for $\gamma\neq 0$ }
In the case of Einstein-Maxwell-dilaton theory,
the Liouville potential leads to the fact that
the solutions are neither asymptotically flat nor de Sitter nor Anti-de Sitter
\cite{chm}. As far as our numerical simulations suggest, this holds also
true for the case of non-abelian gauge fields, since we were not able
to construct asymptotically flat/ de Sitter/ Anti- de Sitter solutions.
However, in a specific limit, namely the embedded abelian case, analytic
solutions are available.

\subsection{The case $H(x)\equiv 1$, $K(x)\equiv 0$}
Setting $H\equiv 1$ and $K\equiv 0$ for all $x$, we find the
following solutions of the system of equations:
\begin{equation}
A(x)=a_0 x^{\frac{\alpha^2}{\gamma^2}} \ , \ \ \psi(x)=\psi_0+\frac{1}{\gamma}\ln(x)
\end{equation}
and
\begin{equation}
N(x)=n_0-n_1 x^{-\frac{(\gamma^2+\alpha^2)}{\gamma^2}} \ \ {\rm with} \ \
n_0=\frac{\gamma^4}{\alpha^2+\gamma^2}\left(e^{2\gamma\psi_0}-
\frac{\Lambda}{2\alpha^2} e^{-2\gamma\psi_0}\right) \ \ . 
\end{equation}
The cosmological constant is given by:
\begin{equation}
\Lambda=2\alpha^2\left(\frac{1}{\alpha^2-\gamma^2}e^{2\gamma\psi_0}-\frac{\gamma^2+\alpha^2}
{\alpha^2-\gamma^2} e^{4\gamma\psi_0}\right) \ .
\end{equation}
This solution has a single event horizon for $n_1 > 0$.
Moreover, it can be seen, that this solution is ill-defined for $\alpha=\gamma$.
Note that these are generalisations of the solutions constructed in \cite{chm}.
For $\alpha=1$, the above solution corresponds to one of the solutions
found in \cite{chm}.
In Fig.3, we show qualitative profiles of the functions for 
the choice of parameters which corresponds to the $5$-dimensional
limit (\ref{5dlimit}). In addition, we choose $\gamma=\psi_0=a_0=n_1=1$.
It is obvious from this figure that the solution has a horizon
(here at $x=x_h \approx 0.655$) and thus represents a black hole.

If we choose instead the limit $\alpha=0$, the function $A(x)$ becomes constant
$=a_0$. The metric function $N(x)=1-n_1 x^{-1}$ in this limit.

\section{Conclusions}
In a previous paper \cite{volkov}, it was shown that a Einstein-Yang-Mills
model in $5$ dimensions can be reduced to an effective $(3+1)$-dimensional
Einstein-Yang-Mills-Higgs-dilaton model under certain symmetry
conditions -spherical symmetry and independence on the coordinate
associated with the $5$th dimension. One of the main results of the present paper
shows that the reduction of a $5$ dimensional de Sitter (dS)/Anti- de Sitter
(AdS)
Einstein-Yang-Mills system to an effective $(3+1)$-dimensional
action (with the same symmetry assumptions as in \cite{volkov}) leads
to a self-interaction of the dilaton field via a Liouville potential.

Previous considerations of an Einstein-Maxwell-dilaton model including
a Liouville potential \cite{chm} have revealed that no asymptotically
flat/ de Sitter/ Anti- de Sitter solutions can be constructed \cite{mitra}.
All our attempts to construct numerically solutions of the non-abelian
counterpart have failed. Thus, we believe that the absence of 
asymptotically flat/ de Sitter/ Anti- de Sitter solutions holds
also true in the case of non-abelian gauge fields. However, considering
the limit of vanishing dilaton coupling, we were able to recover
the AdS gravitating monopoles studied previously \cite{adsmono} and
to produce previously not studied solutions, namely the dS gravitating
monopoles. We show for the latter solutions that they bifurcate
with the branch of dS-Reissner-Nordstr\"om (dSRN) solutions at a critical
value of the gravitational coupling. Finally, considering the limit
$K(x)\equiv 0$ and $H(x)\equiv 1$ for non-vanishing
Liouville potential, we were able to construct generalisations
of the solutions found in \cite{chm}.\\
\\
\\
{\bf Acknowledgements}
Y. B. gratefully acknowledges the Belgian F.N.R.S. for financial support.
B. H. was supported by an EPSRC grant.

\newpage
\begin{figure}
\centering
\epsfysize=12cm
\mbox{\epsffile{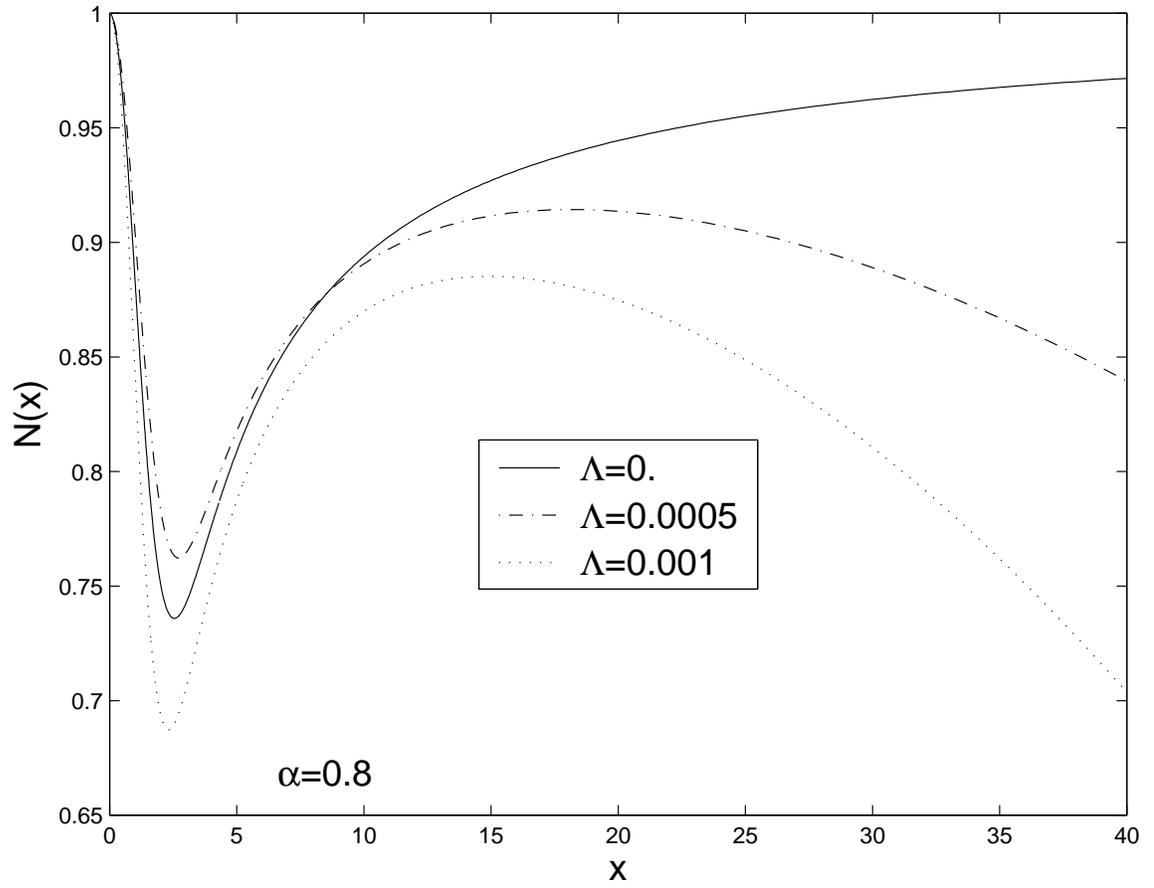}}
\caption{The metric function $N(x)$ is shown for the gravitating monopoles
 with $\alpha=0.8$ and three different choices of the cosmological constant
 $\Lambda$.}
\end{figure}

\newpage
\begin{figure}
\centering
\epsfysize=12cm
\mbox{\epsffile{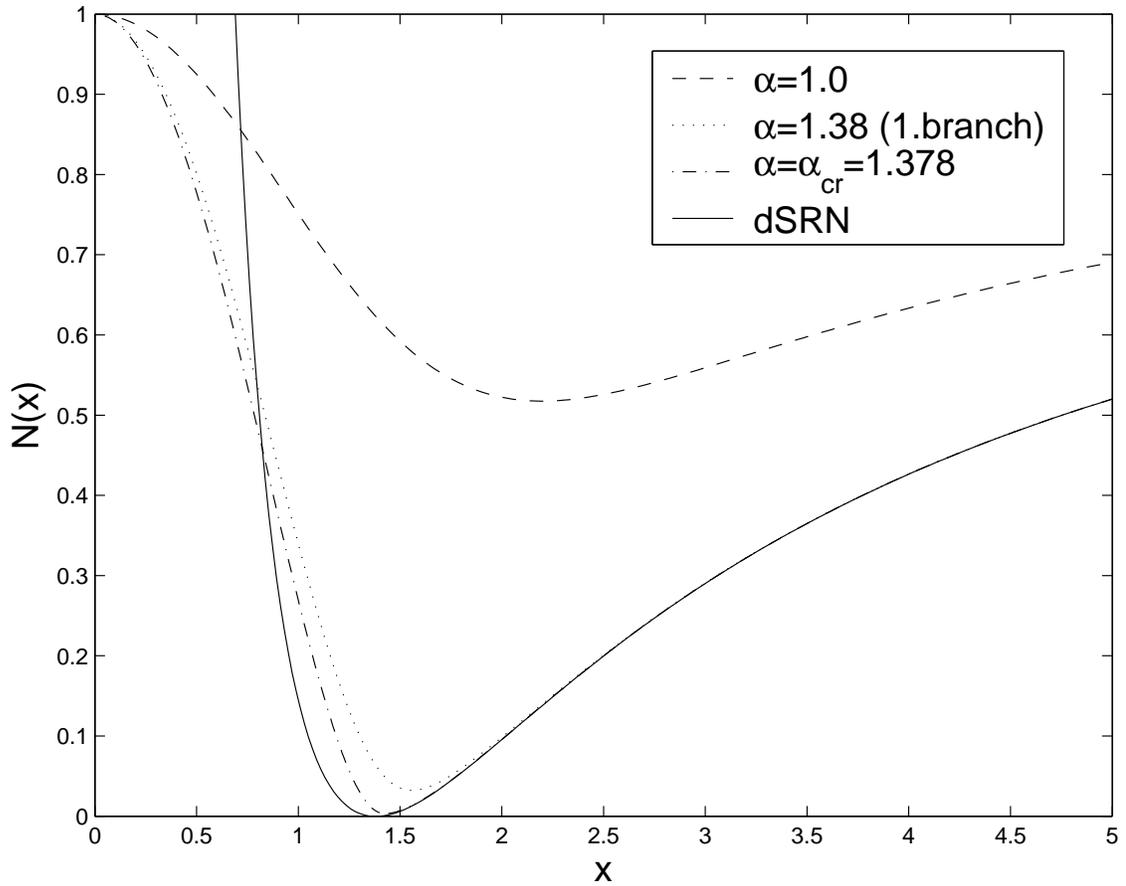}}
\caption{The metric function $N(x)$ of the de Sitter gravitating monopoles
is shown for $\Lambda=0.001$, $\beta=0.1$ and different choices of $\alpha$ including
$\alpha \approx \alpha_{cr}$. For comparison also the corresponding
de Sitter-Reissner Nordstr\"om (dSRN) solution is shown. }
\end{figure}
\newpage
\begin{figure}
\centering
\epsfysize=12cm
\mbox{\epsffile{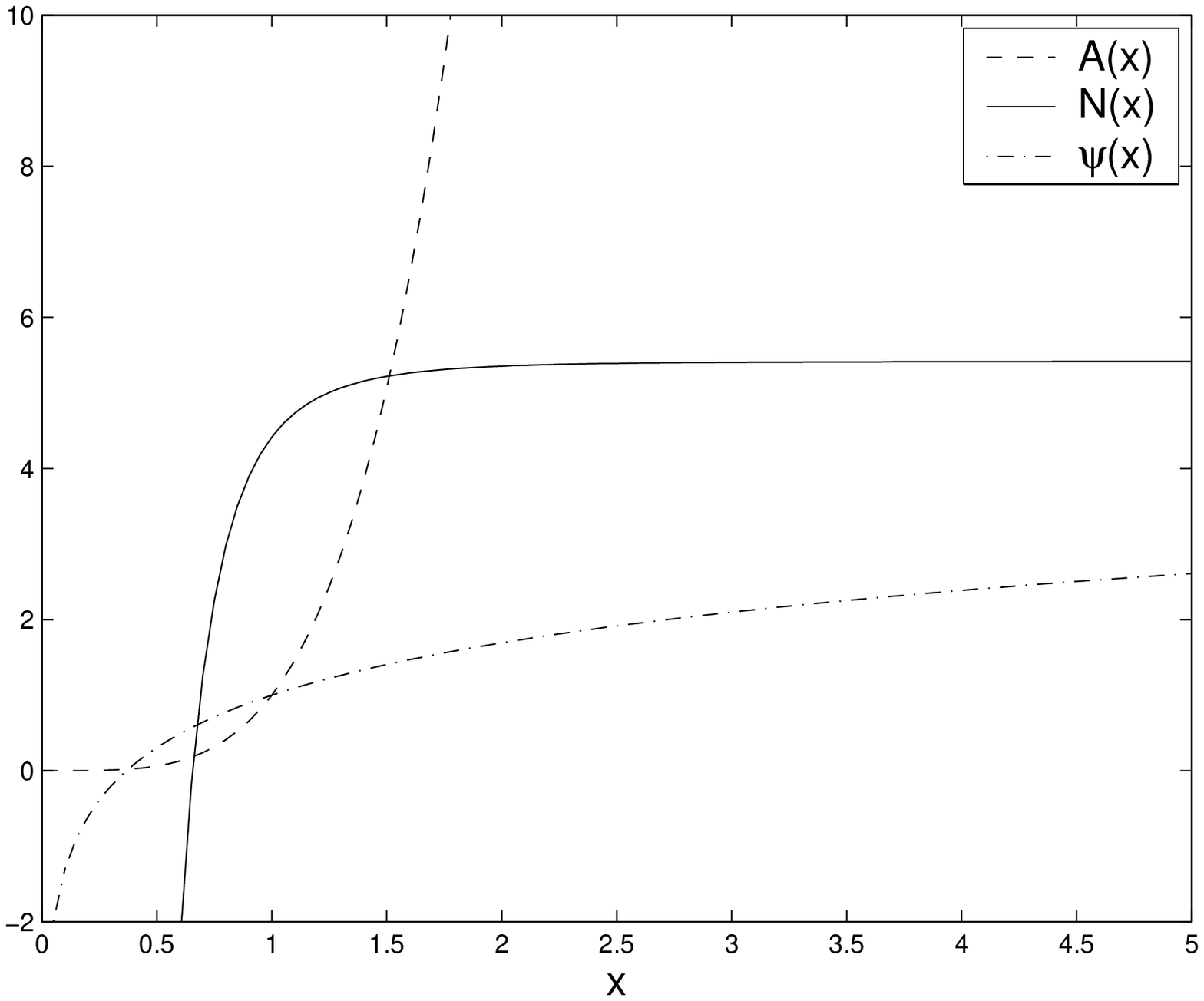}}
\caption{The qualitative profiles of the metric functions $A(x)$ and $N(x)$
and the dilaton function $\psi(x)$ are shown for the $5$-dimensional
parameter limit with $H(x)\equiv 1$ and $K(x)\equiv 0$. }
\end{figure}


\begin{thebibliography}{99}
\bibitem{maeda}
G. Gibbons and K. Maeda, Nucl. Phys. {\bf B298} (1988), 741;
D. Garfinkle, G. Horowitz and A. Strominger, Phys. Rev. {\bf D43} (1991), 371.
\bibitem{chm} K. C. K. Chan, J. H. Horne and R. B. Mann, 
Nucl. Phys. {\bf B447} (1995), 441.
\bibitem{polet} S. Poletti and D. Wiltshire, Phys. Rev. {\bf D50}
(1994) 7260.
\bibitem{mitra} T. Ghosh and P. Mitra, Class. Quantum Grav. {\bf 20} (2003),
1403.
\bibitem{volkov}
M. S. Volkov, Phys. Lett. {\bf B524} (2002), 369.
\bibitem{bh} Y. Brihaye and B. Hartmann , Phys. Lett. {B 534} (2002) 137.
\bibitem{bfm}
 P. Breitenlohner, P. Forgacs and D. Maison,
 Nucl. Phys. {\bf B383} (1992), 357;\\
 P. Breitenlohner, P. Forgacs and D. Maison,
 Nucl. Phys. {\bf B442} (1995), 126.
 \bibitem{weinberg} K. Lee, V. P. Nair and E. J. Weinberg, Phys. Rev. {\bf D45}
 (1992) 2751.
 \bibitem{forgacs1}
P. Forgacs and J. Gyueruesi, Phys. Lett. {\bf B366} (1996), 205.
\bibitem{adsmono} A. R. Lugo, E. F. Moreno and F. A. Shaposnik, 
Phys. Lett. {\bf B 473}
(2000) 35.
\bibitem{thooft}
 G. `t Hooft,  
 Nucl.~Phys.~ {\bf B79} (1974), 276;\\
 A.~M. Polyakov, 
 JETP Lett. {\bf 20} (1974), 194.

\end{thebibliography}
\end{document}